# A Resource Pooling Switch Architecture with High Performance Scheduler

Jiong Du, Fei Hu, Du Xu
Key Laboratory of Optical Fiber Sensing and Communications, Ministry of Education,
University of Electronic Science and Technology of China, Chengdu, P. R. China

*Abstract*—**With the rapid development of network-based services and applications, current network nodes' data plane solutions, which focus on packets forwarding traditionally, are not optimally addressing new requirements such as function flexibility, performance scalability and resource utilization, etc. In this paper, we propose a novel data plane structure called Resource Pooling Switch Architecture (RPSA), which utilizes global shared resource pool to provide different processing functionalities and capacities to different packets/flows. As all network functions are instantiated by the resources connected to the switching fabric and constructed in the form of Service Function Chains (SFC), the traffic patterns changed a lot. We design Balanced Service Capacity based FIRM algorithm (BSC-FIRM) to overcome the deficiencies of classical scheduler in RPSA. Simulation results show that our algorithm outperforms on the aspect of packet switching delay and loss rate compared with FIRM and iSLIP algorithm.**

## I. Introduction

The endless new emerging network-based services and applications need the underlay network evolve accordingly, but so many different traffics are hosted by today's IP network and all serviced in the best-effort manner in part of their end-to-end paths. The design principles of traditional enclosed network nodes, simple and fast, make it focus on the forwarding performance but hardly consider the newly coming requirements.

It is quite easy to point out many deficiencies of the "black box", here we only consider two aspects – inflexibility and poor resource utilization. For the inflexibility, we cannot add new functions required into the box easily as the fast running pipeline structure of line-card implemented by ASIC or NP (network processor) which is not opened to user or third party, and we need many different middle-boxes connected around the nodes to support new functions. For the poor resource utilization, we cannot tailor the processing capacity of the line card to fit different traffic classes as all packets pass all the stages of the pipeline. That means some function modules in the pipeline process the packets, but do no contribution, wasting processing resource absolutely.

To address above problems, P4 (Programming Protocol-Independent Packet Processors) was proposed with a high-level language applied to program the data processing behavior of data plane [2]. With P4, users could program network functionalities directly by coding, compiling and then downloading the microcode to underlying devices to meet requests of packets. However, since packet processing functions supported by data plane become more and more complicated, P4 and its corresponding programmable architecture gradually show limitations in some scenarios, such as highly dynamic flows, new functions' deploy of P4 cannot meet the changing rate. In addition, as Tofino still utilizes pipeline structure with programmable functions of its action stages, it also suffers from the problem of poor resource utilization.

On the other hand, ideas of Network Function Virtualization (NFV) [3-6] and SFC provide important inspiration for the flexible customization and function expansion of the data plane. Based on general computing platform, such as x86 servers and Linux, Virtualized Network Functions (VNFs) are instances running on virtual machines or containers with specified purposes. Eliminating the restrictions of enclosed "black-box", new network functions are able to be developed quickly and deployed flexibly with the help of NFV, and the corresponding resources would also be allocated on-demand. On the users' viewpoint, both an end-to-end routing path and a set of ordered VNFs deployed along the path construct a SFC, which could satisfy the user requests instead of all functions being integrated in one node. But for pursuing multiplex gain which leads to high resource utilization, different flows may share one VNF instance belonging to different SFCs. As the burst nature of network traffic, the VNF distributed across nodes may be working under light load or heavy load, which causes the processing delay to change frequently and drastically. This makes it unpractical to predict and control the end-to-end flow performances, such as delay, jitter and loss rate, which usually influence the end users' quality of experience (QoE) a lot [7].

In this paper, we propose RPSA architecture to solve the problems mentioned above. Based on SFCs constructed from the sharing resource pool, this architecture will provide flexible and customized network functions for data plane of network nodes. We design a classifier in the line card of RPSA to divide the packets into different types by rule matching and encapsulates the packets that need to be processed through the sharing resource pool to identify the service function paths. The scheduler performs one or more packet scheduling on all packets to realize network function processing and switching forwarding in RPSA.

To find out an efficient packet scheduling in RPSA, as different types of flows and the corresponding SFCs processed in sharing resource pool leading the traffic to be varied and non-uniform, we design an novel algorithm called BSC-FIRM in this paper, which defines a new metric called Service Capacity (SC) to quantize the status of each queue during scheduling and optimizes the iterative scheduling of FIRM[8] based on SC to achieve high-performance packet scheduling.

The remaining of this paper is organized as follows: Section Ⅱ introduces the RPSA architecture. Section Ⅲ describes the design of BSC-FIRM algorithm and section Ⅳ presents experiment and analysis of simulation results. Finally, section Ⅴ concludes the paper with generalizing our mainly contributions and perspectives to future work.

## II. NODE ARCHITECTURE

In traditional network node, any packet arriving at it goes through all the functions in the pipeline of line card whether functions are necessary or not, such as ACL, Load Balance, NAT, etc. After that, the packet is scheduled from input to output port of the switch fabric, then leaves the node's ports.

Instead of forwarding to the output ports determined by its destination address directly, such MAC or IP, the packet finds its SFC firstly in RPSA if needed, which is done by the Classifier, and then is handled with the corresponding SFC located in Network Function Pool (NFP). Compared with the classical architecture, the RPSA separates specified functions from line cards' pipeline and allocates them in NFP to provide flexible processing for packets. The block diagram of RPSA is shown in figure 1.

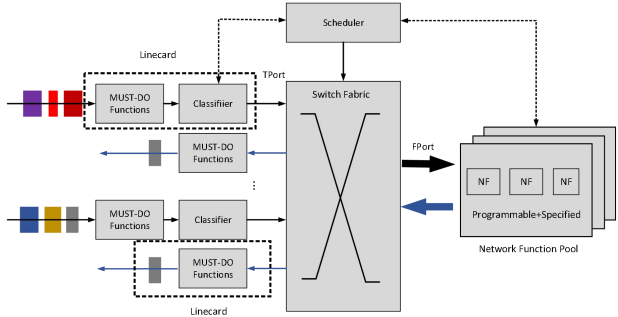

Fig.1 Block Diagram of RPSA

In RPSA, we adopt the well-known VOQ based switching structure as the fabric. The ports of the fabric can be distinguished as two types, one is Transport Port (TPort) and the other is Function Port (FPort).

TPorts connect to the line card and support user traffic IO. But here the line card is simplified to only support MUST-DO Functions and Classifier. Composed of Ingress Pipeline and Egress Pipeline, MUST-DO Functions include some essential network functions in line card, like integrity and legality verification to filter out invalid packets. Classifier is responsible for dividing packets into different types corresponding to different SFCs by rule matching.

FPorts connect to the NFP and each of them is a standard physical port connecting to x86 server. The number of FPorts is in accordance to the number of computer nodes in NFP. By centralizing specified VNFs running on virtual machines or containers through visualizing x86 servers NFP makes it easier to provide various and dynamic network functions for switch data plane.

Scheduler performs one or more scheduling on all packets, including packets that switch directly and tagged packets that need to be processed through NFP, from input to output port with high speed. But as TPorts and FPorts connect to different modules and different flows correspond to different processings, Scheduler in RPSA need special design.

In RPSA, NFP, Classifier and Scheduler are core modules which we will introduce in detail.

### A. NFP

As mentioned above, Each computing nodes in NFP has several Network Function Instances (NFIs), which are VNFs running on servers through NFV. Each NFI is connected by software switch or control modules within nodes. By centralizing those NFIs, NFP can manage all network functions easily and provide a range of flexible and customized services based on users' requests in the form of SFC.

Certainly, there exist some problems if NFIs are placed and managed unproperly in NFP. For example, some SFC consists of several NFIs that may run on many different servers, which results in a large increase in packets processing delay and ineffective transmissions of links between switch fabric and NFP that causes a decline in system throughput. So an optimal placement in NFP should make a SFC goes through as few servers as possible. In this paper, we adopt MFMTP algorithm in [9] to solve the problem of network function placement.

### B. Classifier

The Classifier separates the input packets according to pre-defined rules, which are based on statistics and prediction to the traffic. The whole rules can be described as Forward Information Base (FIB) and NFI State Table in detail. FlB mainly indicates information of SFC and output TPort; NFI State Table indicates information of NFI Index, Queue Length and Function Description. So the ordinary packets, which only need to be forwarded to the output ports, are tagged with TPorts' number based on FIB; the peculiar packets, which need specific functions process, are tagged with the FPorts' number and ordered NFI index based on deployment strategy.

### C. Scheduler

The Scheduler in RPSA is different from the traditional one. Besides scheduling packets from input to output port, it also provides NFI management and corresponding Service Function Path (SFP) management. The working mechanism of Scheduler is as shown in figure 2.

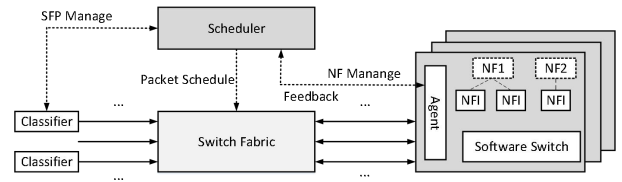

Fig.2. Mechanism of Scheduler

SFP management means that Scheduler assigns SFP for packets according to current state of NFI and SFC. As each server in NFP has an Agent to interact with Scheduler through out-of-band channel, Scheduler could obtain real-time state and processing ability of each NFI in the whole system through messages sent from Agent. Due to this mechanism, Scheduler could assign suitable SFP to packets by instructing Classifier to tag corresponding SFC header. With real-time feedback, it also provides sufficient information to implement management of NFI and dynamic expansion.

Packet scheduling refers to that scheduler solves competition for switch fabric by matching input and output ports to the most extent, so as to make rational use of network resources, reduce latency and improve throughput. In RPSA, the goal of packet scheduling is to achieve efficient transmission between input ports and output ports and good performance under different traffic models. We will discuss the scheduling algorithm in detail in chapter 3.

In conclusion, The whole packet processing in RPSA is as following. Firstly, packet goes through all MUST-DO functions when it arrives at the ingress line card. Then, Classifier matches packet header with FIB updated by Scheduler to get corresponding SFP. If the SFP is empty and the next hop is not Drop, the packet queues the VOQ of destination output TPort and waits to be scheduled. If the SFP is not empty and the next hop is not Drop, Classifier tags the packet with unique header in accordance to SFP, which leads packet to get specified network functions handled. After that, packet queues in VOQ of assigned output FPort when packet gets all specific network function processed in NFP. Finally, Scheduler matches packet header with FIB again to get output TPort pointed by next hop and schedule the packet to leave the switch.

Compared with traditional switch architecture and P4, RPSA outperforms in at least following aspects:
- Processing capacity and rules for packets can be flexibly and dynamically changed. As specified network functions are provided by NFP, the composition and strategy could adjust according to users' requests.
- Resource utilization can be improved. By deep programming for network functions, efficient recycling and reusing of existing resources, RPSA could obviously improve resource utilization.

In addition to bringing advanced and flexible network functions to data plane, different from classical SFC, RPSA also focuses on the performance of the whole system, which includes network function placement that has been addressed in literature [9], and the scheduling problem brought by traffic classification and NFP that will be discussed in next section. By studying these issues, we will realize the completely closed loop of management and scheduling of switch nodes. As far as we know, the current studies only take some sub-problems into account [10-13].

### III. PACKET SCHEDULING

Under the condition of multi-user competing for shared network resources, especially for switch fabric, it is unavoidable to lead to contention and even blocking, which will seriously affect the performance of switch. So for each switch architecture, efficient packet scheduling algorithm is a powerful guarantee to ensure the performance of the whole system.

RPSA is no exception and even more complicated in scheduling, because part of flows need to enter NFP and get specified network functions processed. As different flows correspond to different SFPs, the traffic of RPSA always become non-uniform. This problem could be illustrated simply by figure 3, Flow B is scheduled directly from input TPort to output TPort and Flow A gets all network functions processed in one server while Flow C need to go through two servers. Obviously, Flow C spends more time within switch and consumes more link bandwidth compared with Flow A and B.

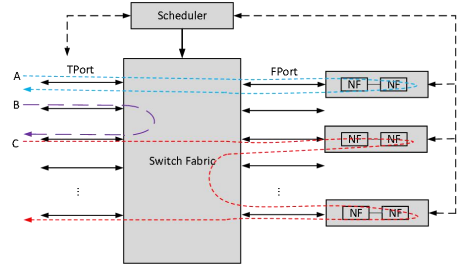

Fig.3. Different SFC processing

So RPSA needs an efficient algorithm to address packet scheduling of non-uniform traffic to ensure efficient transmission between input and output ports so as to maximize the performance of switch. Aiming at input-queued switch, there exist some classic scheduling algorithms like iSLIP[14] and FIRM[8] and many other recent studies. For example, Selective Request Round Robin (SRRR) algorithm expands phases one iteration to four [15]. Literature [16] modulates PIM and iSLIP algorithm to realize two-phase iteration. RR/LQF algorithm adds one more bit to identify whether prior VOQ is empty or not [17]. Although there are a great many packet scheduling algorithms, some of them still need to be optimized, especially in non-uniform traffic model. In this paper, based on iteration scheduling and its representative FIRM algorithm, we propose BSC-FIRM algorithm to solve the problem of non-uniform traffic scheduling in RPSA architecture.

*A. FIRM algorithm*

FIRM is a Maximal Size Matching (MSM) algorithm based on iterative scheduling [8]. It has been demonstrated that the performance of FIRM is slightly better than that of iSLIP [14] under high load conditions. In FIRM algorithm, each input and output port has a strategy pointer *RRI* and *RRO* respectively. Both of them point to highest priority of each port, which could help to match and update strategies. In each iteration, FIRM algorithm adopts three phases of Request - Grant - Accept to complete a maximal size match between the input and output port. It has been proved that FIRM has good performance for uniform traffic. However, when it comes to non-uniform traffic, FIRM still treats and schedules each VOQ equally even though the input rate of each VOQ is different, which leads to longer queuing delay and increasing loss rate.

*B. Service Capacity*

We propose BSC-FIRM algorithm based on FIRM to solve the packet scheduling of RPSA in this paper. In BSC-FIRM, we define Service Capacity (SC) as an important index to quantize the state of each queue in packet scheduling. The SC of each VOQ is co-decided by the length and the latest scheduled time.

Firstly, We define $VOQ_{i,j}$ as VOQ whose input port is $i$ and output port is $j$ and $Len_{i,j}$ as queuing length of $VOQ_{i,j}$. In addition, $T_{i,j}$, described by timeslot, represents the latest time that $VOQ_{i,j}$ is scheduled. The relationship between VOQ and SC is as following:
- If $Len_{i,j}$ is greater than $Len_{i,k}$, the SC value of $VOQ_{i,j}$ is less than that of $VOQ_{i,k}$;
- If $Len_{i,j}$ equals to $Len_{i,k}$ and $T_{i,j}$ is less than $T_{i,k}$, the SC value of $VOQ_{i,j}$ is less than that of $VOQ_{i,k}$;
- If $Len_{i,j}$ equals to $Len_{i,k}$ and $T_{i,j}$ equals to $T_{i,k}$, the SC value of $VOQ_{i,j}$ equals that of $VOQ_{i,k}$.

Figure 4 shows the state of VOQ in the two input ports at a given time. According to the definition of SC metioned above, we get $SC_{i,j} < SC_{i,k}$ and $SC_{x,m} < SC_{x,n}$. For each VOQ, the smaller the value of SC, the more packets accumulated in queue or longer time distance from the latest scheduling is, which is less likely to hold new packets. Therefore, scheduling algorithm should give priority to VOQ with smaller value of SC, which leads SC to be more balanced.

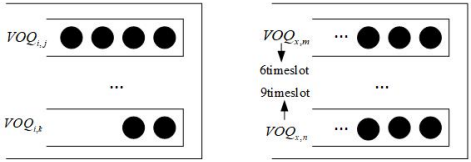

Fig.4. Comparison between SC

### C. BSC-FIRM Algorithm

The basic idea of BSC-FIRM algorithm is that the strategy pointer *RRI* of each input port points to the output port corresponding to the VOQ with the minimum SC in the first iteration of each scheduling. Namely, the input port gives priority to the signal sent by the output port in accordance to the VOQ with minimum SC. If there is more than one VOQs with minimum SC, random selection strategy would be adopted to balance SC of all VOQs. Therefore, Each iteration of BSC-FIRM algorithm performs following three phases:
- Request: All unmatched input ports send Request to unmatched output ports corresponding to non-empty VOQ. If it is the first iteration of the scheduling, strategy pointer *RRI* of the input port would point to the output port corresponding to the VOQ with the minimum SC, and if there is more than one VOQ with minimum SC, *RRI* would select one of them randomly.
- Grant: The unmatched output ports find and send Grant signal to input port that is closest to Grant pointer *RRO* according to round-robin strategy and, at the same time, sends Reject signal to other input ports.
- Accept: After receiving Grant signal, the unmatched input port also sends Accept signal to output port which is closet to pointer *RRI* according to round-robin strategy and try to establish a match. When match successes in the first iteration, *RRI* points to next output port that has set up matching port. In terms of those output port that has sent Grant signal yet not been matched, *RRO* points to input port which has not accepted Grant.

Figure 5 shows an example to illustrate the iteration process of BSC-FIRM algorithm. As shown in figure 5(a) and figure 5(b), pointer *RRI* always points to output port corresponding to VOQ with minimum SC at the beginning of first iteration and all VOQs with minimum SC could get matched after two iterations.

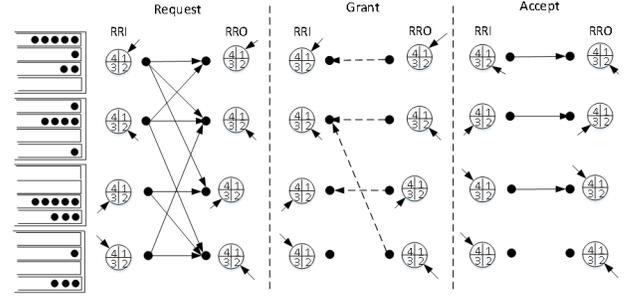

Fig.5(a). Three Phases of The First Iteration

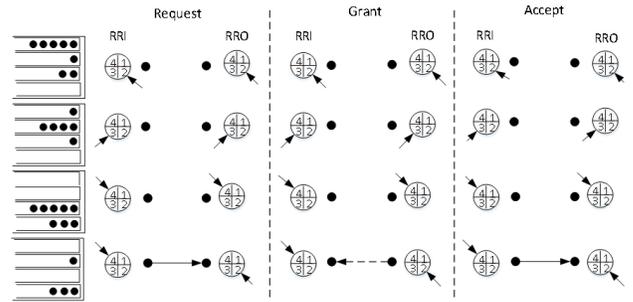

Fig.5(b). Three Phases of The Second Iteration

It has been proved that the time complexity of FIRM algorithm is $O(N^2)$ and $N$ is the number of ports of switch. Compared with FIRM algorithm, BSC-FIRM adds strategy selection of input port pointer based on comparing SC at Request phase of first iteration, which could be finished within $O(\lg N)$. As updating strategies at Grant and Accept are in line with those of FIRM algorithm, the overall time complexity of BSC-FIRM algorithm is $O(N^2 * \lg N)$. Considering the balanced direction each VOQ moves during scheduling, BSC-FIRM algorithm could perform better in aspects of delay and packet loss rate than FIRM algorithm when it is non-uniform traffic. From analysis of time complexity above, we know that the performance improvement attained by BSC-FIRM algorithm is not at too much time cost.

## IV. SIMULATION

### A. Environmental setup

In order to evaluate the performance of BSC-FIRM algorithm, we simulate it under three different traffic models and compare the result with that of FIRM and iSLIP algorithm. In this paper, we adopt 32x32 switch fabric, among which,16 ports are TPort and 16 ports are FPort. The most common SFC proposed in literature [10] including NAT, FW, IDS and VPN is adopted in this paper and the number of deployed NFIs that each NF need is calculated according to the load of NF that includes total traffic arriving at NF, remaining resource of each server and load balance. The network function placement at the granularity of NFI could refer to our another paper [9]. By modulating the total traffic entering the RPSA, namely link load, we could test the performance of three algorithms. The processing delay to cell of each NFI in the simulation is distributed uniformly between slot [1,2]. Classifier uses the source IP address of sending host and UDP source port of packet as the basis to divide traffic. Buffer of each NFI is set to be infinite. The size of cell is 64byte. Each VOQ could accommodate 500 cells at most and use independent buffer. And we set 5 iterations in three algorithms.

In the experiment, we consider three classic traffic models--uniform, burst and hotspot. Considering the characteristics of RPSA architecture, we simulate and compare BSC-FIRM with FIRM and iSLIP [8] under the three traffic models as following:

- Uniform-Uniform: Flows of each sub-SFC and direct switching distribute uniformly among sending hosts. Destination addresses also distribute uniformly among receiving hosts. Each packets arrives according to Bernoulli process.
- Uniform-Hotsport: Flows of each sub-SFC distribute uniformly among sending hosts. Destination addresses distribute uniformly among receiving hosts. Flows of direct switching adopts hotspot traffic model. For input port $i$, the possibility that packets are sent to hotspot host $(i+N/2) \mod N$ is 0.5, so as with the possibility of being sent to other non-hotspot hosts.
- Butst-Burst: Flows of each sub-SFC and direct switching distribute uniformly among sending hosts. Destination addresses also distribute uniformly among receiving hosts. Each packets arrives according to the burst process (ON-OFF). Namely, the destination addresses of all packets of the same burst is the same and destination addresses of different burst obey uniform distribution.

Following two indexes are used to measure the performance of algorithms:

- Delay -- In this simulation, delay refers to queuing delay, because only this kind of delay is decided by packet scheduling algorithm. Suppose that the total number of cells switched in given time is $N$, and the time that cell queues in VOQ is $D_i$. Then average delay could be described as:

$$average\_delay = \sum_{i=1}^{N} D_i$$

- Packet Loss Rate -- When VOQ buffer has been exhausted, the newly arrived cells will be dropped. Suppose that sending hosts send $N$ cells and M cells of them are discarded. Then the packet loss rate is:

$$drop\_rate = \frac{M}{N}$$

### B. Simulation Results and analysis

Figure 6 depicts the performance of three algorithms under Uniform-Uniform traffic model. As shown, there is no packet loss in this model for three algorithm and no big gap between them until load rate increases up to 0.95, where average delay of BSC-FIRM decreases 10.9% and 10.5% compared with iSLIP algorithm and FIRM algorithm respectively. This is because Uniform-Uniform traffic is the most ideal model in RPSA architecture and three algorithms could implement scheduling within the allowable range of the VOQ buffer. While with the increase of load rate, the influence of partial non-uniformity caused by the SFC request on the scheduling process is beginning to appear, under which condition, BSC-FIRM performs better in average delay.

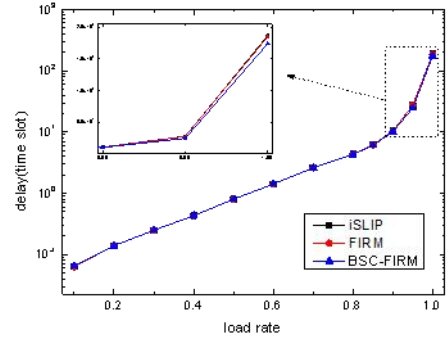

Fig.6. Performance Comparison Under Uniform-Uniform Traffic

The performance of three algorithms under Uniform-Hotspot traffic model is shown in figure 7. From it we could get that when the load rate is greater than 0.8, especially between 0.825 and 0.9, BSC-FIRM has greater improvement than iSLIP and FIRM algorithm in delay and packet loss rate. In terms of load rate where packet loss appears, as shown in figure 7(b), BSC-FIRM algorithm makes an increase of 7.5% compared with iSLIP and FIRM and also decreases 83.7% and 81.9% respectively in the aspect of packet loss rate. This is because BSC-FIRM algorithm gives matching priority to VOQ with minimum SC, leading to packets sent to hotspot host get scheduled in time, which could effectively decrease queuing delay and packet loss rate.

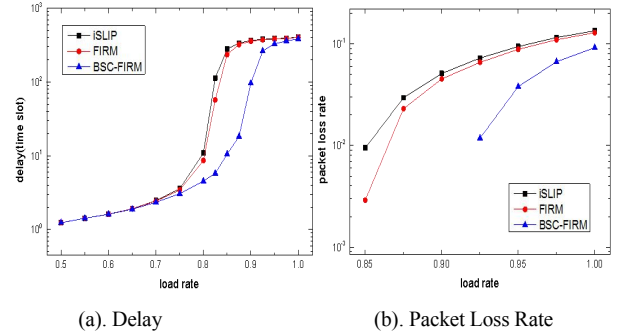

(a). Delay          (b). Packet Loss Rate

Fig.7. Performance Comparison Under Uniform-Hotspot Traffic

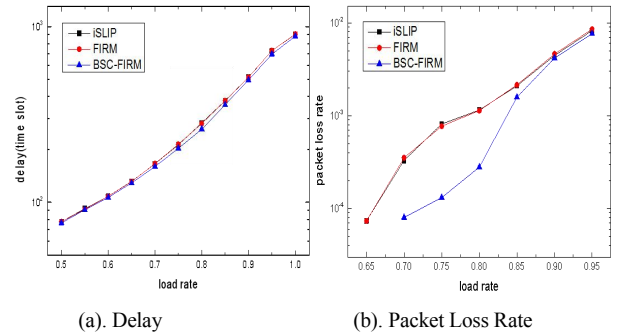

(a). Delay          (b). Packet Loss Rate

Fig.8. Performance Comparison Under Burst-Burst Traffic

Figure 8 shows the performance of three algorithms under Burst-Burst traffic model. From the two pictures, we know that

BSC-FIRM still performs better in both indexes. Comparing Figure 8(a) and 8(b), we know that BSC-FIRM has more improvement in the aspect of packet loss rate than average queuing delay. When the load rates are 0.8 and 0.85, compared with FIRM and iSLIP, the delay of BSC-FIRM algorithm decreases by 8.1% and 5.5% respectively. while at the same load, packet loss rates decrease by 75.3%, and 26.5% respectively. This is because when traffic is in the state of ON, packets with the same destination address constantly arriving at one input TPort of RPSA architecture leads to insufficiency of VOQ buffer, overflow of queue and further increase of packet loss.

From analyzing the simulation results we could conclude that for different traffic models in RPSA architecture, BSC-FIRM algorithm performs better than iSLIP and FIRM algorithm in average queuing delay and packet loss rate during packet scheduling. Therefore, the BSC-FIRM algorithm proposed in this chapter has relatively great potential in applications.

## V. CONCLUSION

In this paper, first of all, we design a switch architecture called RPSA that separates specified functions from line card and provides flexible processing to packets in the form of NFP. This architecture is highly flexible to realize a more intelligent data plane and could assign network functions and resources according to users' requests. Then, considering the switching performance of this architecture, we propose BSC-FIRM algorithm to address packet scheduling under different traffic models. Simulation results show that BSC-FIRM outperforms iSLIP and FIRM algorithm in aspect of average delay and packet loss rate.

As perspectives for future work, First of all, we plan to realize collaboration and control between multiple switch nodes. For one thing, in large-scale networking, such as 3-Clos, network function processing between different switch nodes could collaborate and been flexibly controlled to provide resource reservation and path configure in advance for some traffic with high priority. For another, we plan to introduce the logically centralized control framework like SDN to our architecture so that the whole system will have two levels of data plane capabilities to facilitate the flexible and various bussiness orchestration. Secondly, due to the simulation of this paper, applying RPSA architecture and BSC-FIRM algorithm in practice is also a main task in future research.

ACKNOWLEDGMENT

This work was supported in part by the National 863 Plan Program under Grant 2015AA016102.